\documentclass[letterpaper, 10 pt, conference]{ieeeconf} 
\IEEEoverridecommandlockouts                              
\usepackage{amsmath}
\usepackage{flushend}
\usepackage{graphicx} 
\usepackage{hyperref}
\hypersetup{
	colorlinks=true,
	linkcolor=black,
	citecolor=black,
	filecolor=black,
	urlcolor=black,
}
\usepackage{cleveref}
\crefformat{footnote}{#2\footnotemark[#1]#3}

\usepackage{microtype} 

\usepackage{xcolor}
\definecolor{bg}{rgb}{0.95,0.95,0.97}

\newcommand{\bmatrixx}[1]{\begin{bmatrix}#1\end{bmatrix}}

\title{Linear ADRC is equivalent to PID with set-point weighting and measurement filter}
\author{Fredrik Bagge Carlson, JuliaHub Inc.}
\date{November 2024}

\begin{document}

\maketitle

\begin{abstract}
We show that linear Active Disturbance-Rejection Control (ADRC) tuned using the "bandwidth method" is equivalent to PI(D) control with set-point weighting and a lowpass filter on the measurement signal. We also provide simple expressions that make it possible to implement linear ADRC for first and second-order systems using commonplace two degree-of-freedom PID implementations. The expressions are equivalent to ADRC in the response from measurements, and a slight approximation in the response from references.
\end{abstract}

\section{Introduction}
Active Disturbance-Rejection Control (ADRC) is often touted as a promising improvement over the PID controller as the goto tool for practitioners. In \cite{herbst2013simulative}, the author point out that ADRC, in particular linear ADRC, is a new name for the well-established method of using state feedback with state estimates provided by an observer augmented with a disturbance model. Linear ADRC makes the particular choice of an integrating disturbance model acting on the plant input. 

The author of \cite{herbst2013simulative} compares linear ADRC, tuned using the "bandwidth method" \cite{gao2003scaling}, to PI and PID controllers tuned to a particular response to a reference step. The conclusion that is drawn is that ADRC is generally more robust and performs overall better than the PI and PID controllers.

In this paper, we show that linear ADRC tuned with the bandwidth method is in fact equivalent to a commonplace implementation of a PID controller. For \emph{first-order systems}, the controller response to measurements is equivalent to a PI controller with a first-order filter on the measurement signal, and the response to references is a filtered PID controller. With very minor approximation error, the complete two degree-of-freedom\footnote{Two inputs $r$ and $y$ with different response to each input.} controller can be represented as a single PI controller with set-point weighting and a first-order lowpass filter on the measurement, a very convenient conclusion considering how commonplace this particular implementation of a PI controller is. For \emph{second-order systems}, with a similar minor approximation (in reference response only, no approximation is made in the response from measurements), the linear ADRC controller is equivalent to a PID controller with set-point weighting and a second-order lowpass filter on measurements. Once again, a very convenient conclusion considering the ubiquity of this particular implementation.

Linear ADRC can thus be "implemented" by simply tuning a PI(D) controller with gains derived from the ADRC tuning parameters, and we can view linear ADRC as new PID "tuning rule".

\section{Technical Details}
We will follow the notation used in \cite{herbst2013simulative} for first-order systems, that considers an ADRC controller on the following form
\begin{align}
   \bmatrixx{\dot{\hat{x}}_1 \\ \dot{\hat{x}}_2} &= 
   		\bmatrixx{-l_1 & 1 \\ -l_2 & 0} \bmatrixx{\hat{x}_1 \\ \hat{x}_2} + 
		\bmatrixx{b_0 \\ 0} u + \bmatrixx{l_1 \\ l_2} y \\
	u &= \dfrac{-K_P(r-y) - \hat{f}}{b_0} = \dfrac{-K_P(r-\hat{x}_1) - \hat{x}_2}{b_0}
\end{align}
where the state-space system represents the disturbance observer, $r$ is the reference signal, $y$ is the measurement signal, $u$ is the control signal, and $\hat{f}$ is the estimated disturbance input. The tuning parameters are the observer parameters $l_1$, $l_2$, and $K_P$, and these are set using rules that require the user to pick a desired settling time $T_s$ and the \emph{characteristic gain} of the plant, $b_0$. The plant is assumed to be on the form
$$P(s) = \dfrac{K}{Ts+1}$$
which gives $b_0 = K/T$. The parameters as dictated by the bandwidth tuning rule are selected as 


\begin{align}
	K_P = \dfrac{4}{T_s} \quad  l_1 = -2gK_P \quad	l_2 = g^2K_P^2
\end{align}
where $g$ is a multiplier that determines how much faster the observer poles are compared to the closed-loop poles associated with the plant. In the experiment section for first-order systems in \cite{herbst2013simulative}, $T_s = 1$, $g = 10, K = T = 1$ are used.

For deeper technical details and the formulation for second-order systems, we refer the reader to \cite{herbst2013simulative}.

\subsection{First-order linear ADRC and PID}
\subsubsection{Transfer functions}

When represented as a transfer function, the ADRC controller detailed above can be written as\footnote{We omit the parameter $b_0$ here for brevity. It scales the controller gain proportionally and is thus straightforward to incorporate by scaling the resulting controller by $1/b_0$. The when the equivalent PI(D) parameters are stated, we include $b_0$ for completeness.}



\begin{align}
	C_{ADRC}(s) = \bmatrixx{C_r(s) & -C_y(s)}\\
	C_r(s) = \dfrac{4T_{s}^2s^2 + 32T_{s}gs + 64g^2}{s(T_{s}^3s + T_{s}^2(8g + 4))} \\
	C_y(s) = \dfrac{32T_{s}gs + (16T_{s} + 64)g^2}{s(T_{s}^3s + T_{s}^2(8g + 4))}
\end{align}
 Compare $C_y(s)$ to the transfer function of a PI controller with lowpass filter (PI+F)
$$C_{PIF} = \big(k_p + k_i/s \big)\dfrac{1}{T_f s + 1} = \dfrac{k_p s + k_i}{s(T_f s + 1)}$$
it becomes clear that we can pick $k_p, k_i, T_f$ so that the PI controller becomes identical to the feedback part of the ADRC controller, $C_y$ (including the above omitted parameter $b_0$):
\begin{align}
	k_p &= \dfrac{4 g^2 + 8 g}{b_0 T_s(2 g + 1)} \label{eq:kp}\\
	k_i &= \dfrac{16 g^2}{b_0 T_s^2(2g + 1)} \label{eq:ki}\\
	T_f &= \dfrac{T_s}{8 g + 4} \label{eq:Tf}
\end{align}

The feedforward part, $C_r$, can similarly be shown to be equivalent to a first-order filtered PID controller, but more interestingly, we can approximate $C_r$ as $-C_y$ with set-point weighting on the proportional term, but without the lowpass filter. To do this, we introduce the 2DOF PI controller $K_{ry}$ with set-point weight $b$, defined by the transfer function between $r,y$ and $u$ in the expression
$$K_{ry}(s) : u = k_p (br - y) + k_i(r-y)\frac{1}{s}$$
with the set-point weight $b = K_P / (b_0 k_p)$ where $K_P$ is the proportional gain in the ADRC controller. $k_p$ is the proportional gain in the equivalent PID controller obtained by matching $C_y$ to $C_{PIF}$. With this choice of $b$, the feedforward part of the 2DOF PI controller will have the following low and high-frequency asymptotes:
\begin{align}
	|s| &\to 0:  & \quad sC_r(s) &\to \dfrac{16g^2}{T_{s}^2(2g + 1)} = k_i  \\
	|s| &\to 0:  & \quad sK_{ry}(s) &\to k_i
\end{align}

where we can identify that $K_{ry} \to C_r$ as $|s| \to 0$.

In the other direction, we have
\begin{align}
	|s| &\to \infty:  & \quad C_r(s) &\to \dfrac{4}{T_s} = K_P\\
	|s| &\to \infty:  & \quad K_{ry}(s) &\to b k_p = \frac{K_P}{k_p} = K_P
\end{align}

We thus have the same gain from $r$ to $u$ through $K_{ry}$ as we have through the ADRC controller $C_r$ as $|s| \to 0$ and as $|s| \to \infty$, with a small discrepancy around the crossover frequency.\footnote{Due to the difference in the proportional term in the } Bode curves demonstrating this will be shown for the example provided in \cite{herbst2013simulative} in \cref{sec:numerical}.

\subsubsection{State-space realization of equivalent PIF controller}
For convenience, we also include a statespace realization of the proposed equivalent 2DOF PIF controller (with the approximation of $C_r$):
\begin{align*}
	\dot{x} &= \left[\begin{matrix}0 & - \frac{k_i}{T_f}\\0 & - \frac{1}{T_f}\end{matrix}\right]x + \left[\begin{matrix}k_i & 0\\0 & 1\end{matrix}\right]\bmatrixx{r \\ y} \\
	u &= \left[\begin{matrix}1 & - \frac{k_p}{T_f}\end{matrix}\right]x + \left[\begin{matrix}b k_p & 0\end{matrix}\right]\bmatrixx{r \\ y}
\end{align*}
with inputs $[r, y]$ and output $u$.

\subsection{Second-Order Linear ADRC}
In a similar fashion we can establish equivalence between the second-order linear ADRC controller and a PID controller with set-point weighting and a second-order lowpass filter on the measurement signal, given by the transfer function between $r,y$ and $u$ in the expression:
\begin{align}
	K_{PIDF}(s) &: u = k_p (br - y_f) + (r-y_f)\frac{k_i}{s} - y_fk_d s \\
	y_f &= \dfrac{1}{T_f^2s^2 + 2T_f d s + 1} y	
\end{align}
where $d$ is a relative damping parameter of the filter and $T_f$ is the time constant of the filter. Note that the reference $r$ does not appear in the derivative part, the set-point weight in the derivative term is zero, while $b$ generally is not.

The expressions for $k_p, k_i, k_d, T_f, d$ are more complicated than for the first-order case, but can be derived by matching the transfer function of the ADRC controller to the transfer function of the PID controller in a similar fashion. The expressions are provided in the appendix, and the source code used to symbolically solve for them is provided in \cite{repo}.

\section{Numerical Comparisons} \label{sec:numerical}
This section will demonstrate the equivalence between linear ADRC and the proposed equivalent PI(D) controllers numerically, using the same plant model and experiment as the one used in \cite{herbst2013simulative}. The plant model is a first-order system with $T = 1$ and $K = 1$, and the ADRC controller is tuned with $T_s = 1$ and $g = 10$. The parameters for the PI(D) controllers are found using \cref{eq:kp,eq:ki,eq:Tf} and .

The code to reproduce the results is available in \cite{repo}.

\subsection{First-Order System}

We compare three different controllers, the ADRC controller suggested in \cite{herbst2013simulative}, the PI controller suggested by \cite{herbst2013simulative} (called he "suggested PI(D)"), as well as the filtered PI controller with approximation in the reference response suggested in this paper, the "equivalent PI(D)" $K_{ry}$.

\begin{figure}[h]
	\centering
	\includegraphics[width=0.5\textwidth]{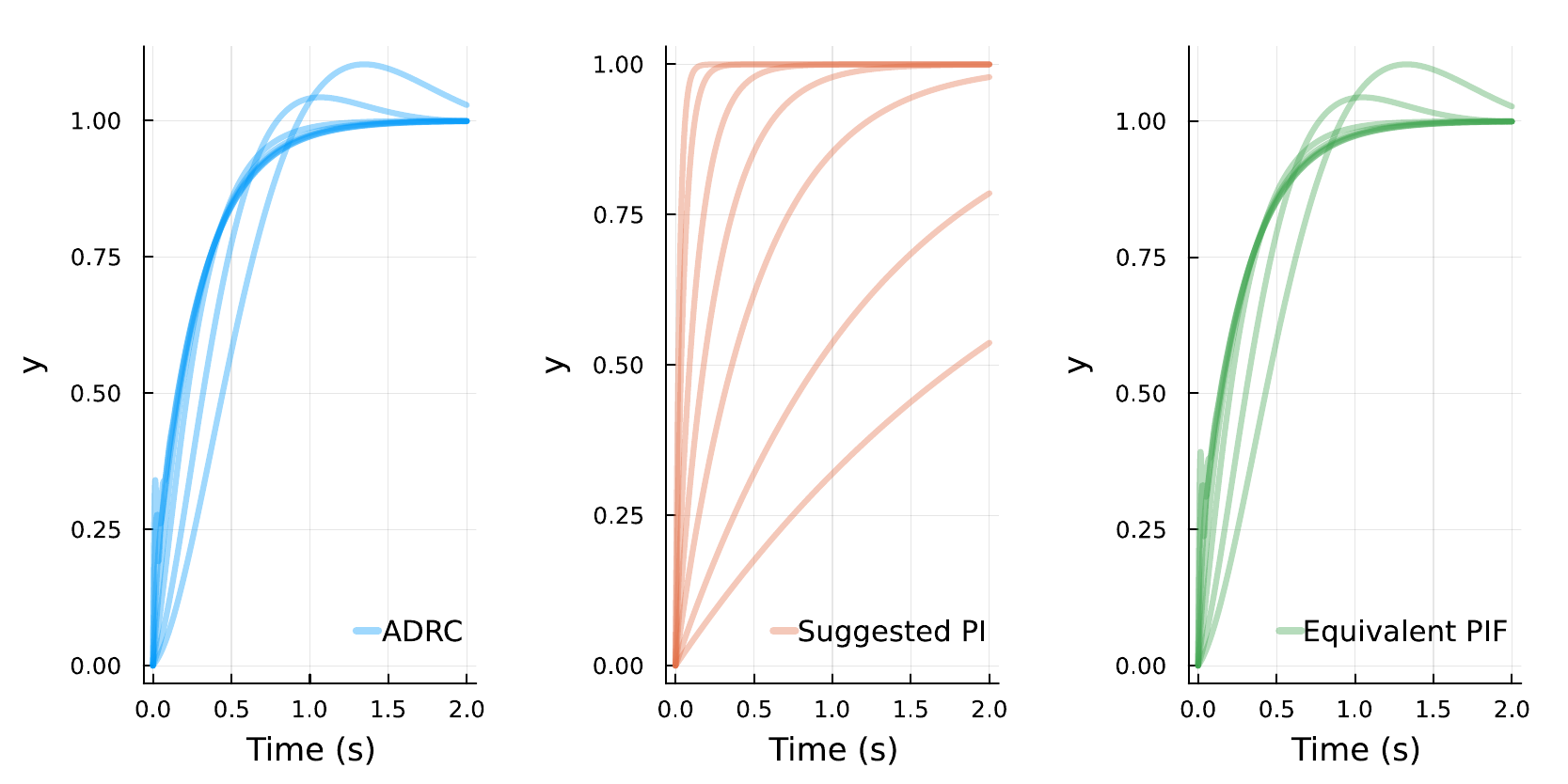}
	\caption{Step response from $r$ to $y$  of the closed loop with different values of the plant gain $K$.}
	\label{fig:first_order_K}
\end{figure}

We start by reproducing the first experiment from \cite{herbst2013simulative} where the parameters of the plant, $T,K$, are varied to simulate model uncertainty. First, we let $K$ take values in the set $\{0.1, 0.2, 0.5, 1, 2, 5, 10\}$. The result is shown in \cref{fig:first_order_K}. As is evident from the figure, the PID controller suggested by \cite{herbst2013simulative} performs much worse than the ADRC controller as $K$ deviates from the nominal value of 1. The "equivalent PID" proposed in this paper performs more or less identically to the ADRC controller, despite the slight approximation. We will look further at this approximation soon, but first we will vary the plant parameter $T$ as well.

In the next experiment, we let $T$ take values in the set $\{0.1, 0.2, 0.5, 1, 2, 5, 10\}$:
\begin{figure}[h]
	\centering
	\includegraphics[width=0.5\textwidth]{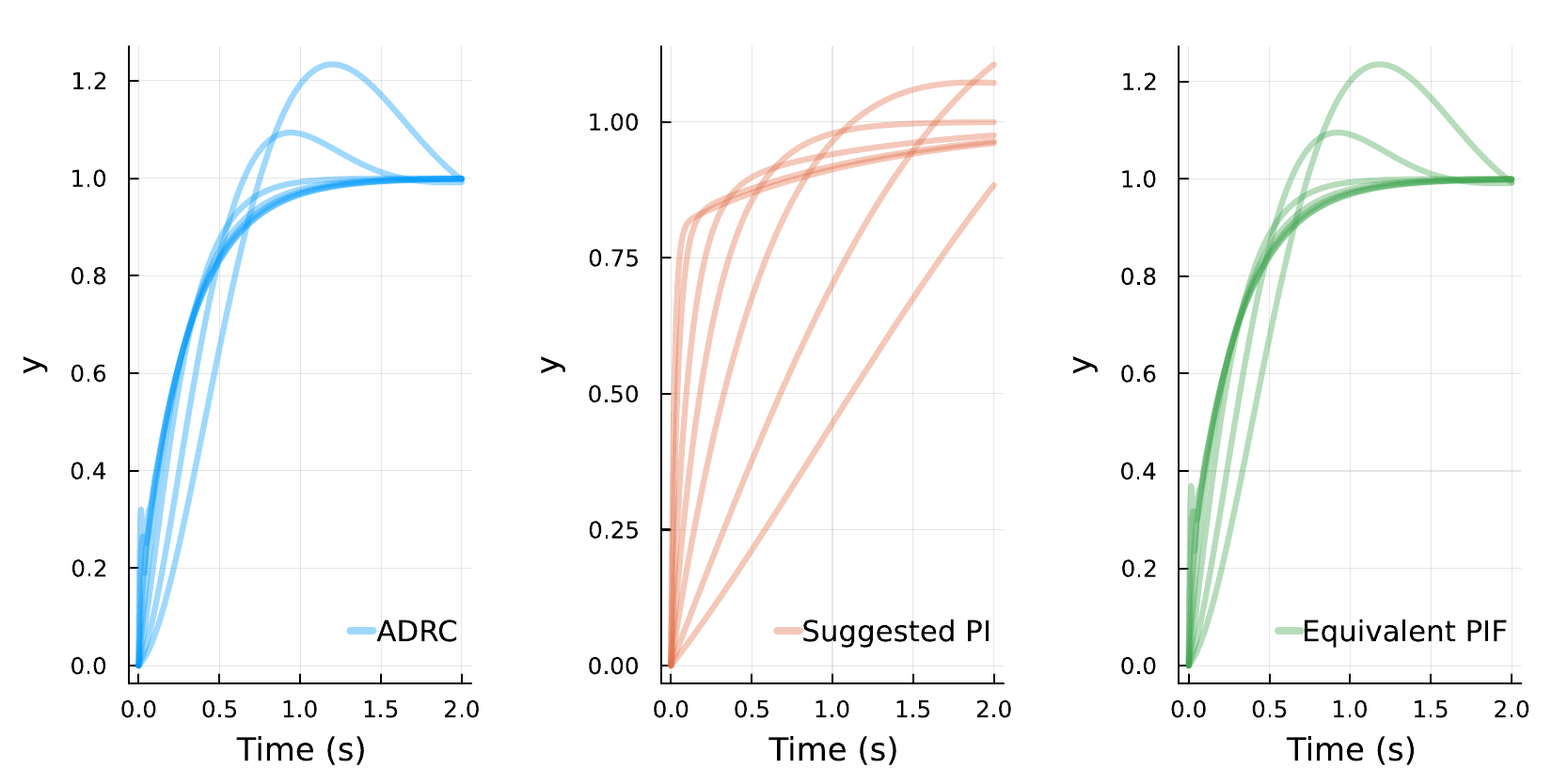}
	\caption{Step response from $r$ to $y$  of the closed loop with different values of the plant time constant $T$.}
	\label{fig:first_order_T}
\end{figure}
once again, the PI controller tuned in \cite{herbst2013simulative} performs poorly, but the ADRC controller and the equivalent PI controller performs well.

\begin{figure}[h]
	\centering
	\includegraphics[width=0.5\textwidth]{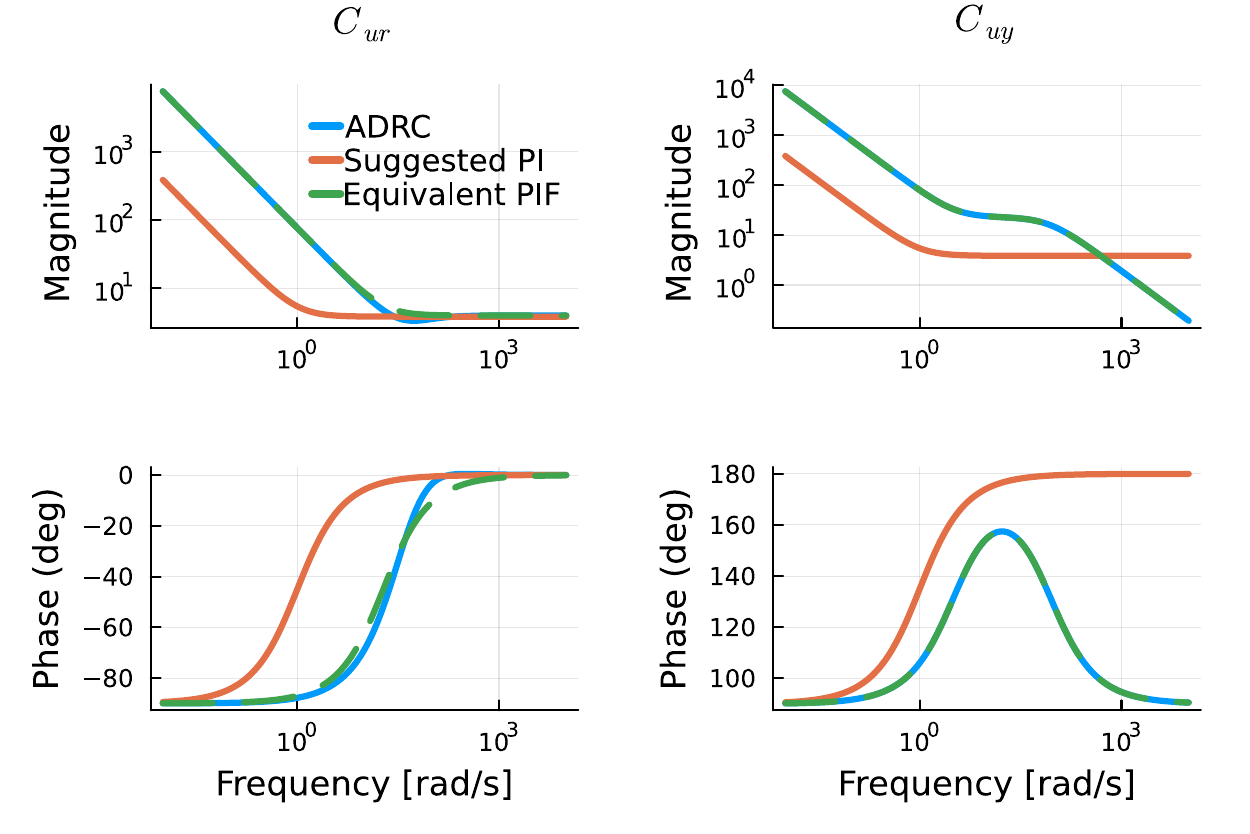}
	\caption{Bode plot of the controllers.}
	\label{fig:first_order_bode_C}
\end{figure}

\begin{figure*}[h]
	\centering
	\includegraphics[width=\textwidth]{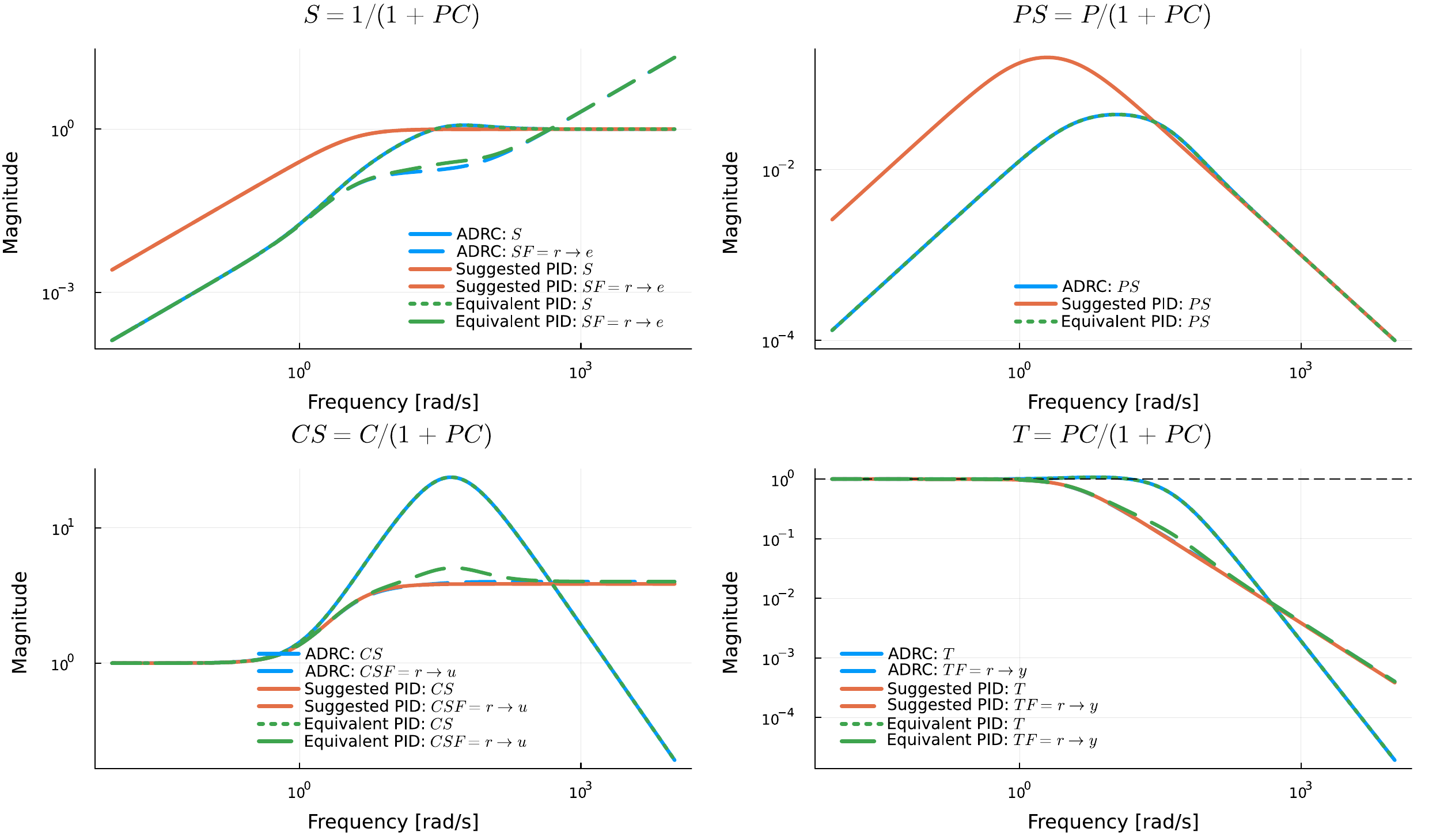}
	\caption{Gang-of-seven plot for the first-order system. The ADRC controller is shown in blue, the PI controller suggested in \cite{herbst2013simulative} is shown in orange, and the equivalent PI controller proposed in this paper is shown in green. The ADRC and the equivalent PI controller are identical in most transfer functions (all solid lines, including only the loop-transfer function) and are hard to distinguish in the plots. The equivalent PI controller does differ slightly from the ADRC controller in the response from references (dashed lines). The plots show the sensitivity function $S = 1/(1+PC)$ (top left), the input disturbance rejection $PS = P/(1+PC)$ (top right), the measurement noise amplification in the control signal $CS = C/(1+PC)$ (bottom left), and the complementary sensitivity function $T = PC/(1+PC)$ (bottom right). Where relevant, the transfer function post multiplied by $F = -C_r / C_y$ is shown as well, indicating the response from references.}
	\label{fig:first_order_7}
\end{figure*}

To understand why the "suggested PI" controller performs poorly, we look at the controller and closed-loop transfer functions, shown in \cref{fig:first_order_bode_C} and \cref{fig:first_order_7} respectively.

The first finding is that the ADRC controller has much higher gain that the suggested PI controller in general, but high-frequency rolloff in the response to measurements. We also see that the approximation made to the response from references in the "equivalent PI controller" is very small (the difference between the blue and the green curve in the left pane). The equivalent PI controller is equivalent to the ADRC transfer function in the response from measurements (right pane), and there is thus no difference in classical measures of robustness since the loop-transfer function is the same.

The (nominal) closed-loop response from references to output, $G_{yr}$, are almost identical between the three controllers, show with dashed lines in the bottom right pane of \cref{fig:first_order_7}. However, the (nominal) closed-loop response from measurements (and measurement noise) to control signal, $G_{uy}$, are not identical, shown in the bottom left pane of \cref{fig:first_order_7}. Here, it is clearly visible that the ADRC controller and the suggested PI controller are very different, where the former has significantly higher gain at intermediate frequencies, but rolloff for high frequencies.

By further looking at the gang-of-seven in \cref{fig:first_order_7}, we can see that the PI controller suggested in \cite{herbst2013simulative} matches the ADRC controller in the closed-loop response from references to output, $TF = PCF/(1 + PC)$, with $F$ defined as $C_r/C_y$. However, the transfer functions involving the loop-transfer function only, the "gang-of-four", paint a different picture. Here, the PI controller is simply using \emph{much less feedback} than the ADRC controller. The suggested PI controller is thus matching the reference step response of the ADRC controller by using higher gain from reference to control signal, corroborated by the bottom left pane of \cref{fig:first_order_7}. Here, we compare the solid orange (and overlapping orange dashed) line here to the blue dashed line. The 2-DOF nature of ADRC and the equivalent PI controller is thus getting away with lower feedforward gain by using higher feedback gain, increasing the robustness to plant variation while keeping the reference step response (close to) identical.

\subsection{Second-Order System}
The second-order system used in the experiment section in \cite{herbst2013simulative} is given by
$$P(s) = \dfrac{1}{T^2s^2 + 2DTs + 1} = \dfrac{1}{s^2 + 2s + 1}$$

\begin{figure}[h]
	\centering
	\includegraphics[width=0.5\textwidth]{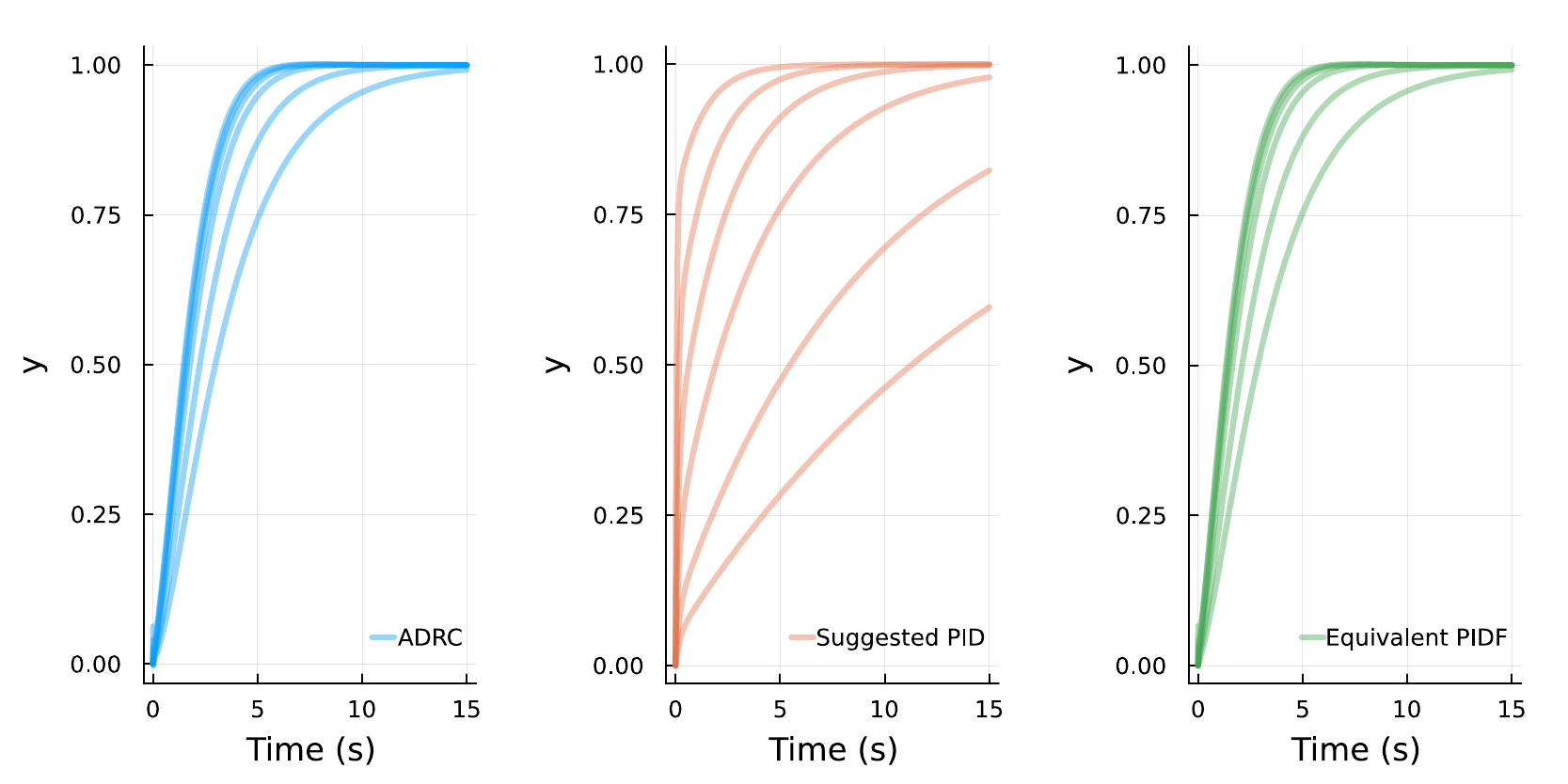}
	\caption{Step response from $r$ to $y$ of the closed loop with different values of the plant gain $K$. Second-order plant.}
	\label{fig:second_order_K}
\end{figure}

\begin{figure}[h]
	\centering
	\includegraphics[width=0.5\textwidth]{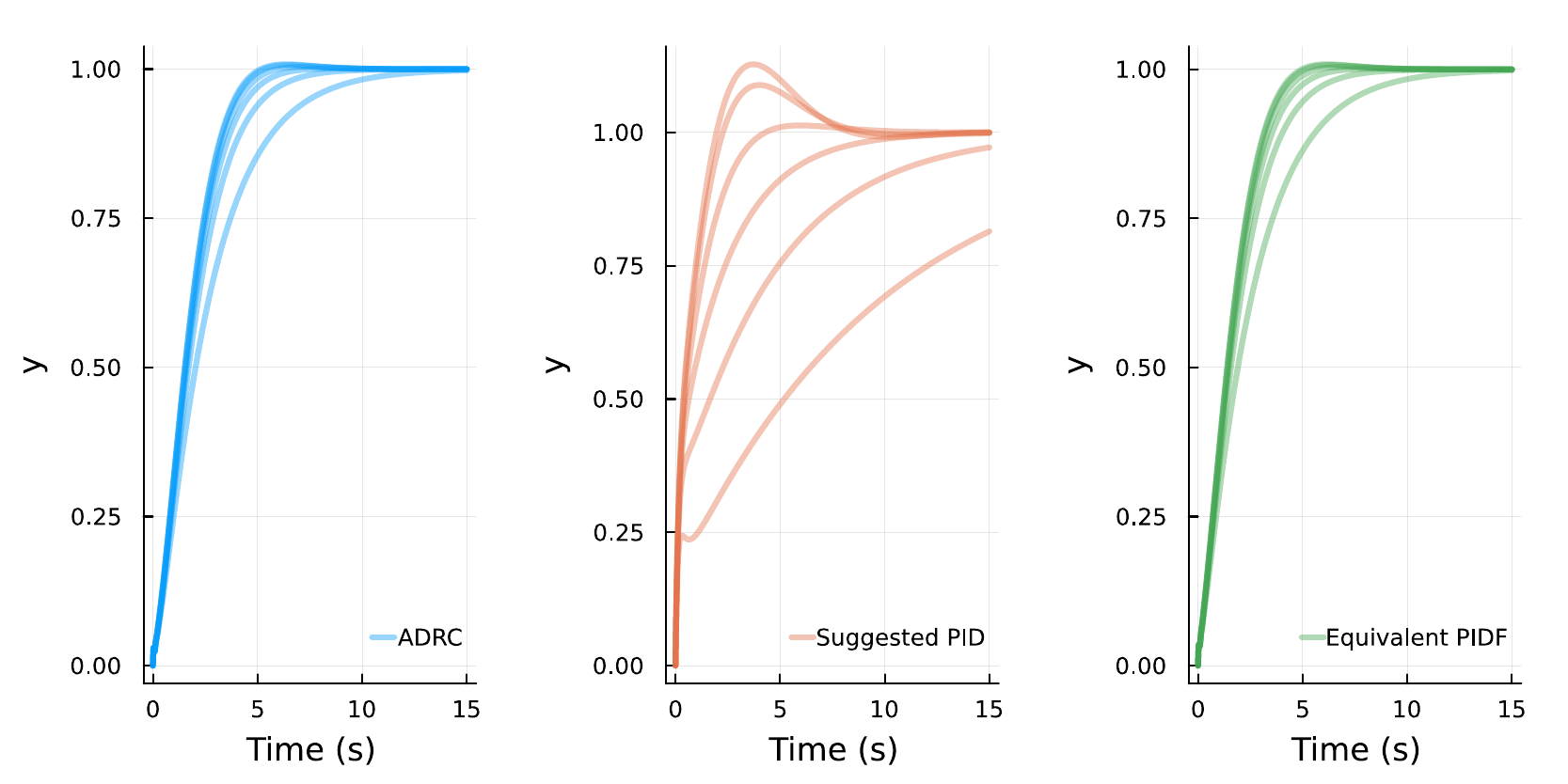}
	\caption{Step response from $r$ to $y$ of the closed loop with different values of the plant time constant $T$. Second-order plant.}
	\label{fig:second_order_T}
\end{figure}

\begin{figure}[h]
	\centering
	\includegraphics[width=0.5\textwidth]{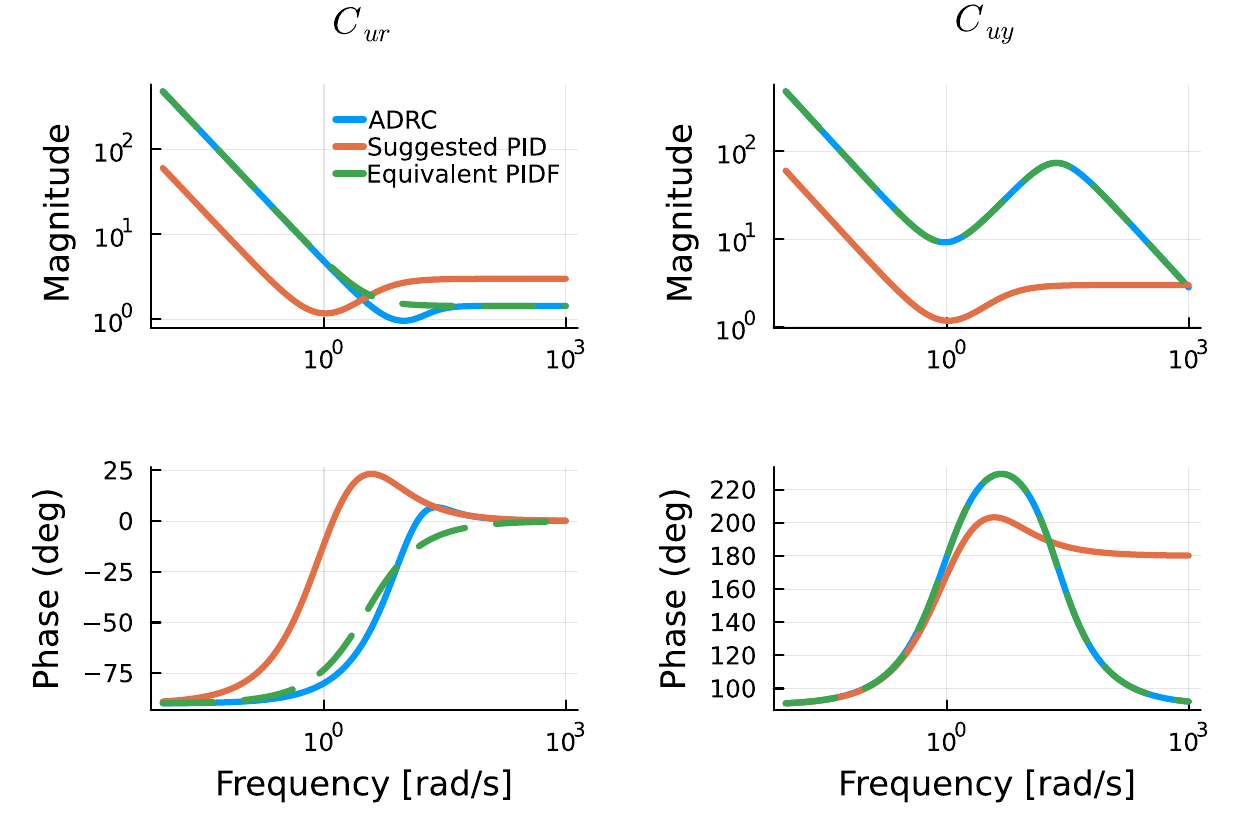}
	\caption{Bode plot of the controllers. Second-order plant.}
	\label{fig:second_order_bode_C}
\end{figure}

\begin{figure*}[h]
	\centering
	\includegraphics[width=\textwidth]{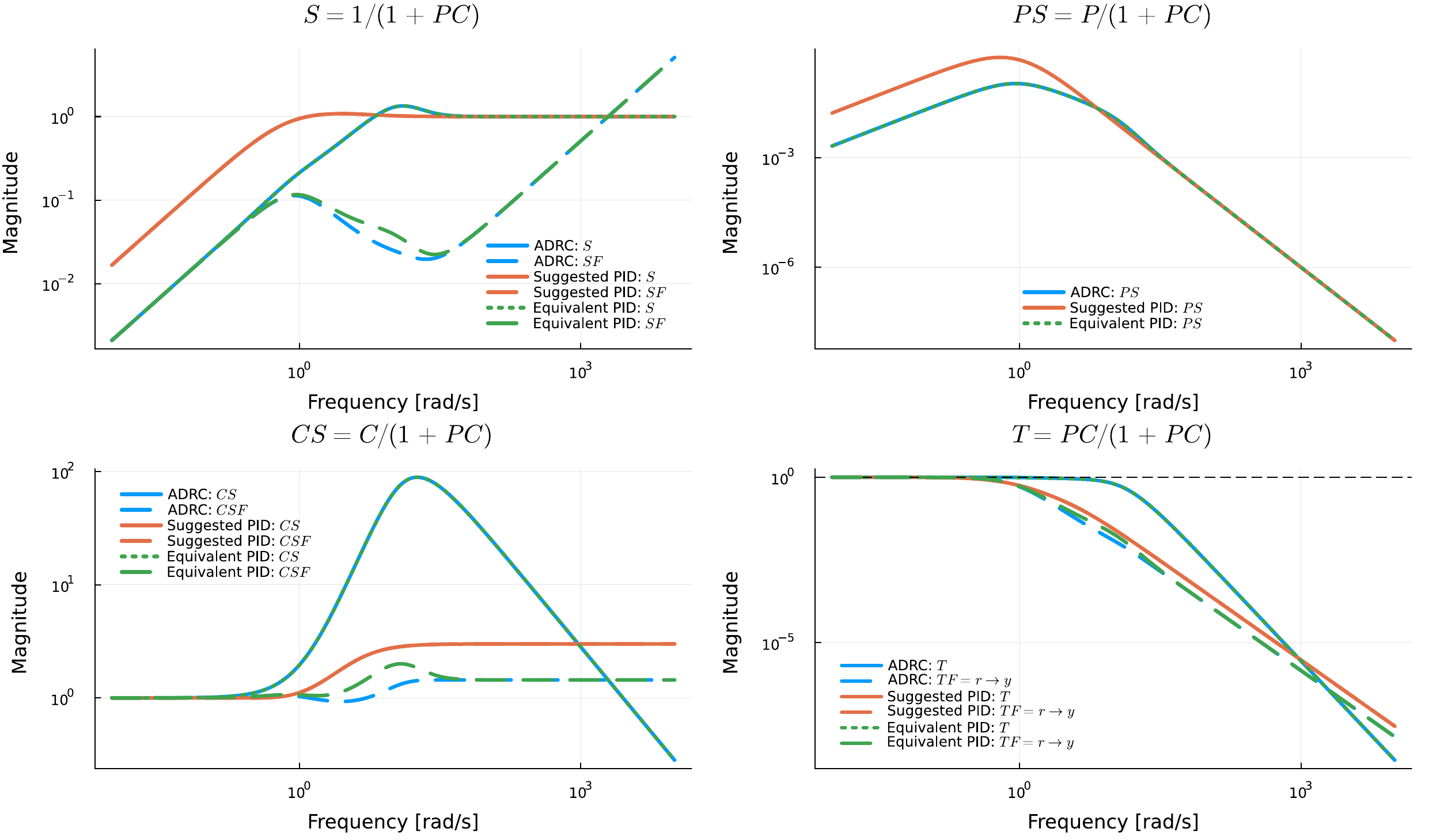}
	\caption{Gang-of-seven plot for the second-order system. The ADRC controller is shown in blue, the PID controller suggested in \cite{herbst2013simulative} is shown in orange, and the equivalent PIDF controller proposed in this paper is shown in green. The ADRC and the equivalent PIDF controller are identical in most transfer functions (solid lines) and are hard to distinguish in the plots. See the legend of \cref{fig:first_order_7} for more details.}
	\label{fig:second_order_7}
\end{figure*}

We show the same figures as for the first-order system
\begin{itemize}
	\item \cref{fig:second_order_K} shows the step response as the value of $K$ varies within the set $K \in \{0.1, 0.2, 0.5, 1, 2, 5\}$.
	\item \cref{fig:second_order_T} shows the step response as the value of $T$ varies within the set $T \in \{0.1, 0.2, 0.5, 1, 2, 5\}$.
	\item \cref{fig:second_order_bode_C} shows the Bode plot of the controllers.
	\item \cref{fig:second_order_7} shows the gang-of-seven for the second-order system.
\end{itemize}
The conclusions are largely the same for second-order systems. The response from measurements is identical between the ADRC controller and the equivalent PID controller. The response to references is slightly different, visible in the bode plot, but not different enough to be clearly visible in the step responses with varying plant parameters.

By looking at the gang-of-seven in \cref{fig:second_order_7}, we once again can see that the PID controller suggested in \cite{herbst2013simulative} matches the ADRC controller in the closed-loop response from references to output, $TF = PCF/(1 + PC)$, with $F$ defined as $C_r/C_y$,  while the "gang of four" paint a different picture.


\section{Discussion}
Given the often reported finding that ADRC controllers are easy to tune and robust w.r.t. plant variations \cite{herbst2013simulative}, the findings in this paper are good news; Not only is the ubiquitous PI(D) controller able to be equally robust and easy to tune, it may in fact be equivalent to the linear ADRC controller. The general concept behind ADRC, that of state feedback from a disturbance observer, is of course much more general. Model-based methods may incorporate more sophisticated knowledge of both the plant and any disturbances acting on it, extending also to nonlinear models. Among model-free and easy-to-use control methods, however, the PI(D) controller remains a strong contender. The fact that the PI(D) controller often comes up short in comparison to ADRC controllers in the literature can often be attributed to either poor tuning or failure to use set-point weighting or filtering, concepts that are commonplace in practical PI(D) controllers. If a filtered 2DOF implementation of a PI(D) controller is used, the ADRC tuning rule can be used to tune also the PID controller if the user so prefers. We end the discussion with a quote from Åström, from one of his many lectures on the principles of feedback:

\begin{quotation}
	It may be strongly misleading to only show properties of a few systems for example the response of the output to command signals. A common omission in many papers and books.\\
	- Karl Johan Åström in "Feedback Fundamentals", 2019
\end{quotation}

\section{Conclusion}
We have demonstrated the equivalence between linear ADRC with parameters selected using the bandwidth method, and a PI(D) controller with set-point weighting and a lowpass filter on the measurement signal. The equivalence is exact in the response from measurements, and a slight approximation in the response from references (an exact but less elegant equivalent controller exist for first-order systems). 

The resulting PI+F controller will for first-order systems have the familiar form
\begin{align}
u &= k_p (br - y_f) + k_i(r-y_f)\frac{1}{s} \\
y_f &= \dfrac{1}{T_f s + 1} y
\end{align}
and for second-order systems, the PID+F controller will be on the form
\begin{align}
u &= k_p (br - y_f) + (r-y_f)\frac{k_i}{s} - y_fk_d s \\
y_f &= \dfrac{1}{T_f^2s^2 + 2T_f d s + 1} y
\end{align}

This highlights that the linear ADRC tuning method can be used as a PID tuning method and implemented using commonplace PID-controller implementations.

\bibliographystyle{IEEEtran}
\bibliography{references}

\appendix

\section{Second-order linear ADRC as PID}
The transfer functions for the second-order ADRC controller (derived with symbolic computer calculations, reproduced in the software repository \cite{repo}) can be realized by the following PID+F parameters (including $b_0$ and the approximation of $C_r$):
\begin{align}
	k_p &= \dfrac{(72g^3 + 108g^2)}{b_0 (3T_s^2 g^2 + 6T_s^2 g + T_s^2)} \\
	k_i &= \dfrac{216g^3}{b_0 (3T_s^3 g^2 + 6T_s^3 g + T_s^3)} \\
	k_d &= \dfrac{(6g^3 + 36g^2 + 18g)}{b_0 (3T_s g^2 + 6T_s g + T_s)} \\
	T_f &= \frac{T_s}{6\sqrt{3g^2 + 6g + 1}} \\
	d &= \frac{3g + 2}{2\sqrt{3g^2 + 6g + 1}} \\
	b &= 36 / (b_0 T_s^2 k_p)
\end{align}

A state-space realization of the PID+F controller above is given below
\begin{align*}
	\dot{x} &= \left[\begin{matrix}0 & 0 & 1\\1 & 0 & 0\\- \frac{1}{T_{f}^{2}} & 0 & - \frac{2 d}{T_{f}}\end{matrix}\right]x + \left[\begin{matrix}0 & 0\\1 & 0\\0 & - \frac{1}{T_{f}^{2}}\end{matrix}\right]\bmatrixx{r \\ y} \\
	u &= \left[\begin{matrix}k_{p} & k_{i} & k_{d}\end{matrix}\right]x + \left[\begin{matrix}b k_{p} & 0\end{matrix}\right]\bmatrixx{r \\ y}
\end{align*}
with inputs $[r, y]$ and output $u$. The state $x$ in this state-space realization represents
$$ x = \bmatrixx{y_f \\ \int r - y_f dt \\ \dot{y_f}}$$
where $y_f$ is the filtered measurement signal.

\end{document}